# Type-II Band Alignment in the $\beta$-Ga$_2$O$_3$/Rutile GeO$_2$ Heterojunction toward Solar-Blind Photodetection: A first-principles study

D. Q. Fang*

MOE Key Laboratory for Nonequilibrium Synthesis and Modulation of Condensed Matter, School of Physics, Xi'an Jiaotong University, Xi'an 710049, China

Abstract

Semiconductor heterostructures capable of separating photogenerated electrons and holes have a wide range of optoelectronic applications, including photodetectors, solar cells, and photocatalysts. $\beta$-Ga$_2$O$_3$ and rutile GeO$_2$ are both ultrawide-bandgap semiconductors, with bandgaps of 4.85 eV and 4.68 eV, respectively. In this work, we employ first-principles calculations based on density functional theory to investigate the band alignment of the $\beta$-Ga$_2$O$_3$/rutile GeO$_2$ heterojunction and explore the effect of interfacial oxygen vacancy. Calculations using the PBE0 hybrid functional based on an interface model show that a type-II band alignment emerges at the $\beta$-Ga$_2$O$_3$/rutile GeO$_2$ interface, which facilitates the separation of photogenerated carriers. The valence band maximum of $\beta$-Ga$_2$O$_3$ lies 0.38 eV below that of rutile GeO$_2$, and its conduction band minimum lies 0.36 eV below. The presence of interfacial oxygen vacancy in the stable configuration leads to a reduction in the band offset. Our results suggest that the $\beta$-Ga$_2$O$_3$/rutile GeO$_2$ heterojunction holds significant promise for application in strictly solar-blind photodetectors.

*fangdqphy@xjtu.edu.cn

# I. Introduction

Gallium Oxide ($Ga_2O_3$), possessing several polymorphs, i.e., corundum (α), monoclinic (β), defective spinel (γ), orthorhombic (κ), and cubic bixbyite (δ) [1–4], has an ultrawide bandgap ranging from 4.5 eV to 5.3 eV [5], depending on the crystal structure. Among these different polymorphs of $Ga_2O_3$, β-Ga2O3 is the most stable phase under normal conditions of temperature and pressure, which has attracted great interest for use in power electronic devices and solar-blind (200-280 nm) photodetectors due to its high thermal and chemical stabilities, large critical electric field strength of ~ 8 MV/cm, and ultrawide bandgap of ~ 4.8 eV [5].

While effective and controllable n-type doping of $Ga_2O_3$ has been achieved using Si, Ge, and Sn shallow donors [6], stable p-type conductivity of $Ga_2O_3$ remains unattainable. In the design of optoelectronic devices, to circumvent the lack of p-type conductivity of $Ga_2O_3$, one of the feasible avenues is to combine $Ga_2O_3$ with other semiconductors to form p-n type [7,8] or n-n type [9,10] heterojunctions. For instance, high-performance self-powered ultraviolet (UV) photodetectors were fabricated based on a GaN/Sn:$Ga_2O_3$ p-n junction film [7] and a CuI/$Ga_2O_3$ p-n junction film [8] in which photogenerated electron-hole pairs are effectively separated via a high built-in electric field. For solar-blind photodetection applications, however, these semiconductor heterojunction devices exhibit an additional photosensitivity to the non-solar-blind region due to the narrower bandgap of the components. For example, the photodetector based on GaN/Sn:$Ga_2O_3$ heterojunction responds not only to the solar-blind region but also to the broader UV spectrum [7].

An alternative strategy for fabricating solar-blind photodetectors involves utilizing $Ga_2O_3$ phase junctions. Recently, α/β- and κ/β-$Ga_2O_3$ phase junctions [11–13] have been grown in experiments. Lu et al. [13] demonstrated that a photodetector based on κ/β-$Ga_2O_3$ phase junction, which features a type-II band alignment, exhibited approximately three orders of magnitude enhancement in photoresponsivity in comparison to the bare β-$Ga_2O_3$ and κ-$Ga_2O_3$ photodetectors.

Rutile germanium oxide (r-GeO2), possessing an ultra-wide bandgap of 4.68 eV [14], has emerged as a promising material for power electronic and optoelectronic devices.

First-principles calculations [15,16] predict that r-GeO$_2$ can be ambipolarly doped and its electron/hole mobilities at 300 K are 244-377/27-29 cm$^2$V$^{-1}$S$^{-1}$, respectively. The measured thermal conductivity for r-GeO$_2$ is 51 Wm$^{-1}$K$^{-1}$ at 300 K [17], which is about two times higher than the highest value of β-Ga$_2$O$_3$ (11 and 27 Wm$^{-1}$K$^{-1}$ along the [100] and [010] directions, respectively) [18]. Recently, experiments [19–21] reported the growth of r-GeO$_2$ using molecular beam epitaxy, metal-organic chemical vapor deposition (CVD), and mist CVD methods. Given that β-Ga$_2$O$_3$ and r-GeO$_2$ have similar bandgap and small lattice mismatch, it is of great interest to explore the heterojunction formed by them, which could serve as a promising structure for strictly solar-blind photodetectors. In this work, we using first-principle calculations investigate the interfacial property and band offset of the β-Ga$_2$O$_3$/r-GeO$_2$ heterojunction.

## II. Computational details

Our first-principles density functional theory calculations are performed using the plane-wave pseudopotential code Vienna *ab initio* Simulation Package (VASP) [22]. The electron-ion interaction is described by the projector augmented-wave method [23,24]. The Ga 3d and Ge 3d electrons are treated explicitly as valence electrons. Structural optimizations are carried out using the PBEsol exchange-correlation energy functional [25] and a plane-wave cutoff energy of 520 eV. The Brillouin zones of the primitive unit cells for bulks β-Ga$_2$O$_3$ and r-GeO$_2$ are sampled using Γ-centered k-point grids of 9×9×5 and 6×6×9, respectively. A vacuum region of 12 Å in the direction perpendicular to the plane of slab and a Monkhorst-Pack k-point grid [26] of 8×4×1 are employed for slab calculations. All atoms are relaxed until the Hellmann–Feynman forces are less than 0.01 eV Å$^{-1}$. To address the bandgap underestimation problem of the PBEsol functional, additional calculations with the PBE0(α) hybrid functional that includes a fraction of Fock exchange α [27] are performed at PBEsol-relaxed geometries using a reduced cutoff of 400 eV, in order to reduce the computational load.

The (100) surface of β-Ga$_2$O$_3$ has two types of termination [28,29], i.e., (100)A and (100)B. To examine the interface stability, we build two types of interface models, i.e., (100)A β-Ga$_2$O$_3$/(110) r-GeO$_2$ and (100)B β-Ga$_2$O$_3$/(110) r-GeO$_2$, with 114 atoms in the unit cell comprising twelve monolayers of β-Ga$_2$O$_3$ and nine monolayers of r-GeO$_2$, as shown in Fig. 1. The in-plane lattice parameters (ILPs) are fixed to the average of the PBEsol lattice parameters of the two materials and the out-of-plane lattice parameter and atoms' internal coordinates are allowed to relax. The interface energy is calculated as [30]

$$\gamma_{\text{int}} = (E_{A|B} - N_A \mu_A - N_B \mu_B)/S - \gamma_A - \gamma_B, \quad (1)$$

where $E_{A|B}$ is the total energy of the interface slab with area S, $N_A$ and $N_B$ are the number of formula units of phases A and B in the interface slab, respectively, and μ is the bulk energy per formula unit including the elastic energy contribution, if any strain is applied. $\gamma_A$ and $\gamma_B$ are the surface energy of phases A and B, respectively.

Coherent interface structure inevitably incorporates strain in the constituent phases. We compute the strained band offset using one interface and two bulk calculations. In the bulk calculations, the ILPs are set to the same set X as in the interface calculations, and the other lattice parameters and internal coordinates are optimized. Taking the macroscopically averaged electrostatic potential in a bulklike region as a reference level [31], the strained valence band offset (VBO) is calculated as [32]

$$\Delta \varepsilon_{VBM,X}^{A-B,strained} = \Delta \varepsilon_{VBM-\text{Ref},X}^{A} + \Delta \varepsilon_{\text{Ref},X}^{A-B} - \Delta \varepsilon_{VBM-\text{Ref},X}^{B}, \quad (2)$$

where $\Delta \varepsilon_{\text{Ref},X}^{A-B}$ is the difference between the reference levels of phases A and B at the ILP X obtained through the calculation of a lattice-matched interface. $\Delta \varepsilon_{VBM-\text{Ref},X}^{A}$ ($\Delta \varepsilon_{VBM-\text{Ref},X}^{B}$) is the energy difference between the valence band maximum (VBM) and the reference level for bulk phase A (B) at the strained lattice constants, which is obtained via bulk calculations.

We also evaluate the natural (unstrained) band offset to eliminate the effect of strain. The change of reference level due to strain is determined by aligning the vacuum levels of unstrained and strained surfaces. The natural VBO is given as [32,33]

$$\Delta\varepsilon_{VBM}^{A-B,natural} = (\Delta\varepsilon_{VBM-Ref,A}^{A} - \Delta\varepsilon_{Vac-Ref,A}^{A} + \Delta\varepsilon_{Vac-Ref,X}^{A}) + \Delta\varepsilon_{Ref,X}^{A-B}$$
$$-(\Delta\varepsilon_{VBM-Ref,B}^{B} - \Delta\varepsilon_{Vac-Ref,B}^{B} + \Delta\varepsilon_{Vac-Ref,X}^{B}) \quad , \quad (3)$$

where $\Delta\varepsilon_{VBM-Ref,A}^{A}$ is the energy difference between the VBM and the reference level for bulk phase A at its natural PBEsol lattice constants. $\Delta\varepsilon_{Vac-Ref,A}^{A}$ ($\Delta\varepsilon_{Vac-Ref,X}^{A}$) is the energy difference between the vacuum level and the reference level close to the center of the slab for phase A at ILP A (X), which is obtained via surface calculations. $\Delta\varepsilon_{VBM-Ref,B}^{B}$, $\Delta\varepsilon_{Vac-Ref,B}^{B}$, and $\Delta\varepsilon_{Vac-Ref,X}^{B}$ denote similar quantities for phase B. The conduction band offset (CBO) is obtained by adding the difference of the calculated bandgaps of phases A and B to the VBO.

### III. Results and discussion

We first investigate the bulk properties of β-Ga$_2$O$_3$ and r-GeO$_2$. β-Ga$_2$O$_3$ is described by four lattice parameters (Table I), i.e., three lattice vectors and *β*, the angle between the *a* and *c* axes, and r-GeO$_2$ has two lattice parameters, i.e., *a* and *c*. The optimized lattice parameters using the PBEsol functional for β-Ga$_2$O$_3$ and r-GeO$_2$ are in good agreement with experimental values (Table I). As shown in Table II, however, PBEsol severely underestimates the bandgaps, particularly for r-GeO$_2$. Calculations using the PBE0(α) hybrid functionals with α = 0.25 and α =0.285 for β-Ga$_2$O$_3$ and r-GeO$_2$, respectively, show a bandgap close to experimental data.

A. Structural property and interface energy of the β-Ga$_2$O$_3$/r-GeO$_2$ heterojunction

Figures 1(a) and 1(b) show the relaxed atomic structures of the (100)A β-Ga$_2$O$_3$/(110) r-GeO$_2$ and (100)B β-Ga$_2$O$_3$/(110) r-GeO$_2$ interfaces, respectively. Both interfaces have in-plane lattice parameters of *a* = 2.969 Å and *b* = 6.037 Å, with a 90° angle between them. There are modest strains (< 4%) in β-Ga$_2$O$_3$ and r-GeO$_2$, as shown in Table III. At the (100)A β-Ga$_2$O$_3$/(110) r-GeO$_2$ interface, the Ga atoms form bonds with the O atoms of r-GeO$_2$, with a bond length of 1.97 Å, and the Ge atoms form bonds with the O atoms of β-Ga$_2$O$_3$, with a bond length of 1.87 Å, whereas at the (100)B β-Ga$_2$O$_3$/(110) r-GeO$_2$ interface, only the Ga atoms form bonds with the O atoms of r-GeO$_2$, with a bond length of 2.22 Å.

For β-Ga$_2$O$_3$, the (100)A surface has a higher surface energy compared with the (100)B surface [29]. When forming interface with r-GeO$_2$, the interface energy of the (100)A β-Ga$_2$O$_3$/(110) r-GeO$_2$ interface is calculated to be 11 meV Å$^{-2}$, which is considerably lower than that of the (100)B β-Ga$_2$O$_3$/(110) r-GeO$_2$ interface (93 meV Å$^{-2}$). This is consistent with the fact that the interface between (100)A β-Ga$_2$O$_3$ and (110) r-GeO$_2$ is well bonded. In the following, we investigate the electronic property of β-Ga$_2$O$_3$/r-GeO$_2$ heterojunction based on the (100)A β-Ga$_2$O$_3$/(110) r-GeO$_2$ interface model.

B. Band alignment

Figure 2 plots the layer-by-layer projected density of states (PDOS) for the (100)A β-Ga$_2$O$_3$/(110) r-GeO$_2$ interface. No clear interface states exist and the middle regions of β-Ga$_2$O$_3$ and r-GeO$_2$ retain their bulk electronic properties. It is also observed that the valence band top of β-Ga$_2$O$_3$ is lower in energy compared to that of r-GeO$_2$.

For interface calculations using hybrid functional, the fraction of Fock exchange $\alpha_{Interface}$ is set to $(\alpha_{Ga_2O_3} + \alpha_{GeO_2})/2$. The band extrema of each component are aligned to the reference level through bulk calculations based on hybrid functional with a material-specific α. Previous study has demonstrated that this mixed scheme achieves excellent agreement with experimental data for semiconductor-oxide interfaces [34]. Figures 3(a) and 3(b) show the planar-averaged and macroscopically averaged electrostatic potentials along the *c* axis for the (100)A β-Ga$_2$O$_3$/(110) r-GeO$_2$ interface calculated using the PBEsol and PBE0(α) functionals, respectively. The reference level offset $\Delta\varepsilon_{Ref,X}^{A-B}$ is calculated to be 1.54 eV using the PBEsol functional and 1.39 eV using the PBE0(α=0.2675) hybrid functional. To examine the effect of the parameter $\alpha_{Interface}$, we perform a calculation using the PBE0(α=0.25) hybrid functional for interface and observe no change in the value of $\Delta\varepsilon_{Ref,X}^{A-B}$ compared with the PBE0(α=0.2675) hybrid functional.

According to Eq. (2), the strained VBO for the β-Ga$_2$O$_3$/r-GeO$_2$ heterojunction is calculated to be –0.63 eV using the PBEsol functional, which is in accord with the PDOS result, and –0.38 eV using the hybrid functional, as shown in Fig. 4 and Table IV. The PBEsol functional yields inaccurate bandgaps. As such, the PBE0 hybrid functional is employed to calculate the strained CBO, giving a value of –0.36 eV. Our results indicate that the β-$Ga_2O_3$/r-$GeO_2$ heterojunction forms a staggered type-II band alignment, as shown in Fig. 4(b), which facilitates the separation of photogenerated electrons and holes. Thus, the β-$Ga_2O_3$/r-$GeO_2$ heterojunction, combined with the ultrawide bandgaps of its components, is promising for strictly solar-blind photodetectors.

We investigate the natural band offsets according to Eq. (3). Calculations yield a natural VBO of –0.61 eV using the PBEsol functional and –0.41 eV using the PBE0 hybrid functional. The natural CBO, obtained using the PBE0 hybrid functional, is –0.26 eV, as shown in Table IV. The natural and strained band offsets differ by no more than 0.1 eV. From the in-plane strain case [Eq. (2)] to the natural case [Eq. (3)], the VBM position for β-$Ga_2O_3$ remains nearly constant, and for r-$GeO_2$ the shift in its VBM is largely offset by the changes of reference levels. Consequently, the natural VBO is very close to the strained one.

C. Oxygen vacancy at the β-$Ga_2O_3$/r-$GeO_2$ interface

Oxygen vacancy ($V_O$) is a common intrinsic defect in β-$Ga_2O_3$ [35,36] and r-$GeO_2$ [15], which can form during crystal or film growth. To explore the effect of an interfacial oxygen vacancy on the band alignment, we increase the in-plane cell size from 1×1 to 4×1 to allow for more structural degrees of freedom. Here we consider only a neutral vacancy. There are four types of interfacial oxygen ions that are threefold coordinated as shown in Fig.5 (a). We find that an oxygen vacancy at site 1 ($V_{O1}$) is the most stable configuration. The relaxed atomic structure of $V_{O1}$ is shown in Fig. 5(b). The formation energy of $V_{O1}$ under the O-rich condition is calculated to be 3.60 eV, while that of $V_{O2}$, $V_{O3}$, and $V_{O4}$ are higher by 0.71, 1.36, and 0.98 eV, respectively. The formation energy of 3.60 eV is lower than the 4.06 eV calculated for a similar neutral oxygen vacancy in bulk β-$Ga_2O_3$.

When $V_{O1}$ is present at the interface, $\Delta\varepsilon_{\text{Ref},X}^{A-B}$ is determined to be 1.97 eV using the PBEsol calculations, as shown in Fig. 5(c), which is larger by 0.43 eV than the value in the case without oxygen vacancy. For the cases with $V_{O2}$, $V_{O3}$, and $V_{O4}$ at the interface, the value of $\Delta\varepsilon_{\text{Ref},X}^{A-B}$ changes slightly. Consequently, the introduction of $V_{O1}$ significantly reduces the strained VBO, which is calculated to be -0.2 eV, as shown in Table IV. In Figure S1, we align the valence bands of β-$Ga_2O_3$ and r-$GeO_2$ based on the ionization potential (IP). The introduction of oxygen vacancy at the (100)A surface for β-$Ga_2O_3$ leads to a significant decrease in the IP, thus reducing the band offset with respect to r-$GeO_2$, which is consistent with the calculation through the interface model. The interfacial oxygen-vacancy-induced VBO reduction has also been observed in the $TiO_2$/$SrTiO_3$ heterostructure [37]. To preserve the band offset of the β-$Ga_2O_3$/r-$GeO_2$ heterojunction, it is necessary to grow a high-quality interface with a low density of oxygen vacancies.

## IV. Conclusions

In summary, we using first-principles calculations investigate the band alignment of the β-$Ga_2O_3$/rutile $GeO_2$ heterojunction. Based on an interface model with a low interface energy, we find that the valence band maximum of β-$Ga_2O_3$ lies 0.63 eV/0.38 eV below that of rutile $GeO_2$ predicted by the PBEsol functional/PBE0 hybrid functional, respectively, and its conduction band maximum lies 0.36 eV below predicted by the PBE0 hybrid functional. Our results indicate that the strained band offsets exhibit only slight differences from the natural ones, and the presence of interfacial oxygen vacancy in the stable configuration reduces the band offset. A type-II band alignment and the ultrawide bandgaps of the components make the β-$Ga_2O_3$/rutile $GeO_2$ heterojunction promising for application in solar-blind photodetectors.

**ACKNOWLEDGMENTS**

We acknowledge the financial support from the National Natural Science Foundation of China (Grant No. 11604254) and the Natural Science Foundation of Shaanxi

Province (Grant No. 2019JQ-240). We also acknowledge the HPCC Platform of Xi'an Jiaotong University for providing the computing facilities.

Table I. Calculated and experimental lattice parameters (in Å) of β-$Ga_2O_3$ and r-$GeO_2$.

|   | β-$Ga_2O_3$ | | r-$GeO_2$ | |
|---|---|---|---|---|
|   | Calc. | Expt. [38] | Calc. | Expt. [39] |
| $a$ | 12.279 | 12.233 | 4.426 | 4.4066 |
| $b$ | 3.050 | 3.038 | 4.426 | 4.4066 |
| $c$ | 5.815 | 5.807 | 2.887 | 2.8619 |
| $β(°)$ | 103.75 | 103.82 | | |

Table II. Indirect bandgap (in eV) of β-$Ga_2O_3$ and direct bandgap (in eV) of r-$GeO_2$ calculated using the PBEsol and PBE0(α) functionals. The fraction of Fock exchange α for each material is given in parenthesis.

|   | PBEsol | PBE0(α) | Expt. |
|---|---|---|---|
| β-$Ga_2O_3$ | 2.22 | 4.83 (0.25) | 4.85 [40] |
| r-$GeO_2$ | 1.56 | 4.67 (0.285) | 4.68 [14], 4.74 [20] |

Table III. Strain in the in-plane lattice parameters $a$ and $b$ of coherent β-$Ga_2O_3$/r-$GeO_2$ interfaces. Negative values indicate compressive strain and positive values tensile strain.

| β-$Ga_2O_3$ | | r-$GeO_2$ | |
|---|---|---|---|
| $a$ | $b$ | $a$ | $b$ |
| –2.66% | 3.82% | 2.84% | –3.55% |

Table IV. Strained band offsets (in eV) for the β-Ga$_2$O$_3$/r-GeO$_2$ heterojunction without and in the presence of oxygen vacancy at the interface, and natural band offsets for the case without oxygen vacancies at the interface. Negative values indicate that the VBM of β-Ga$_2$O$_3$ is lower than that of r-GeO$_2$.

|  | VBO(PBEsol) | VBO(PBE0) | CBO(PBE0) |
|---|---|---|---|
| Strained | –0.63 | –0.38 | –0.36 |
| Natural | –0.61 | –0.41 | –0.26 |
| V$_{O1}$ | –0.20 | | |
| V$_{O2}$ | –0.64 | | |
| V$_{O3}$ | –0.68 | | |
| V$_{O4}$ | –0.61 | | |

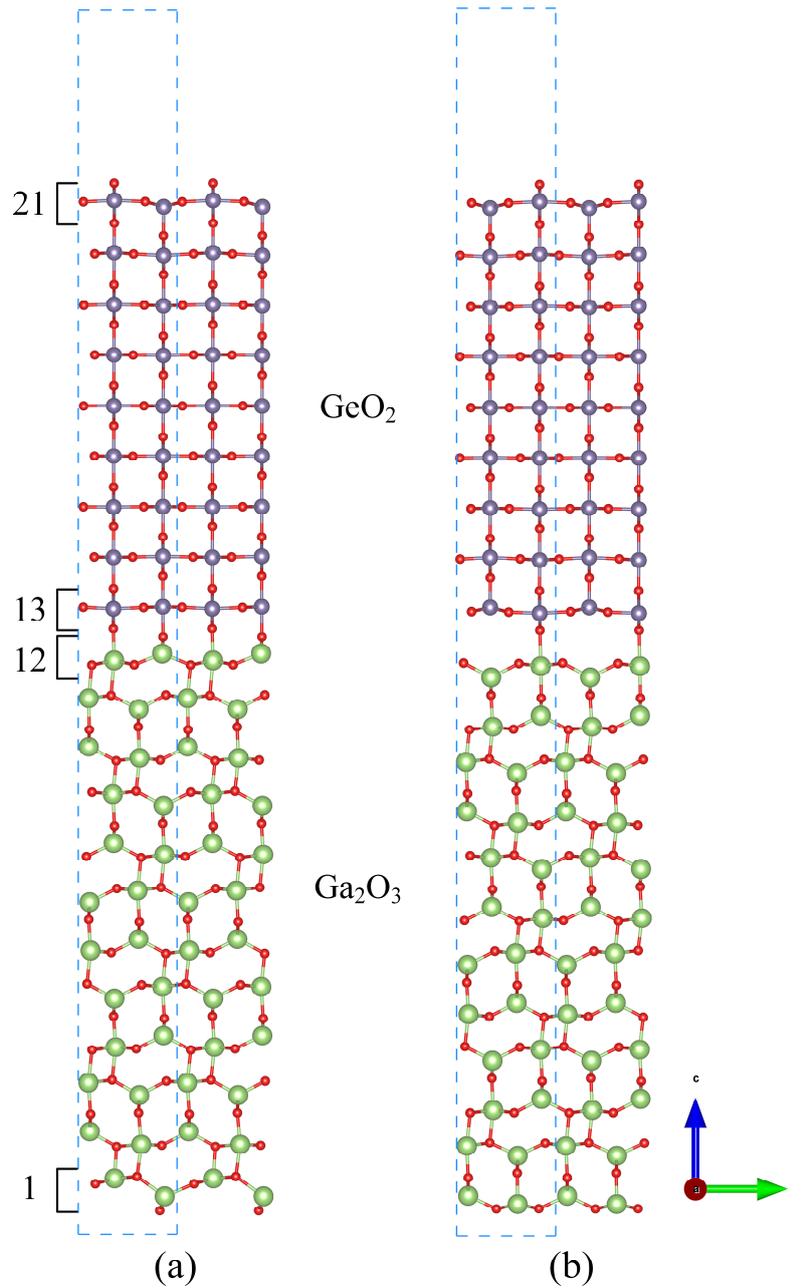

Fig. 1 Optimized atomic structures for the β-Ga$_2$O$_3$/r-GeO$_2$ interfaces: (a) (100)A β-Ga$_2$O$_3$/(110) r-GeO$_2$ interface and (b) (100)B β-Ga$_2$O$_3$/(110) r-GeO$_2$ interface. Ga, Ge and O atoms are represented by green, gray and red balls, respectively. The unit cell of slab is denoted by dashed lines. The top and bottom atomic layers of each material are denoted by numbers.

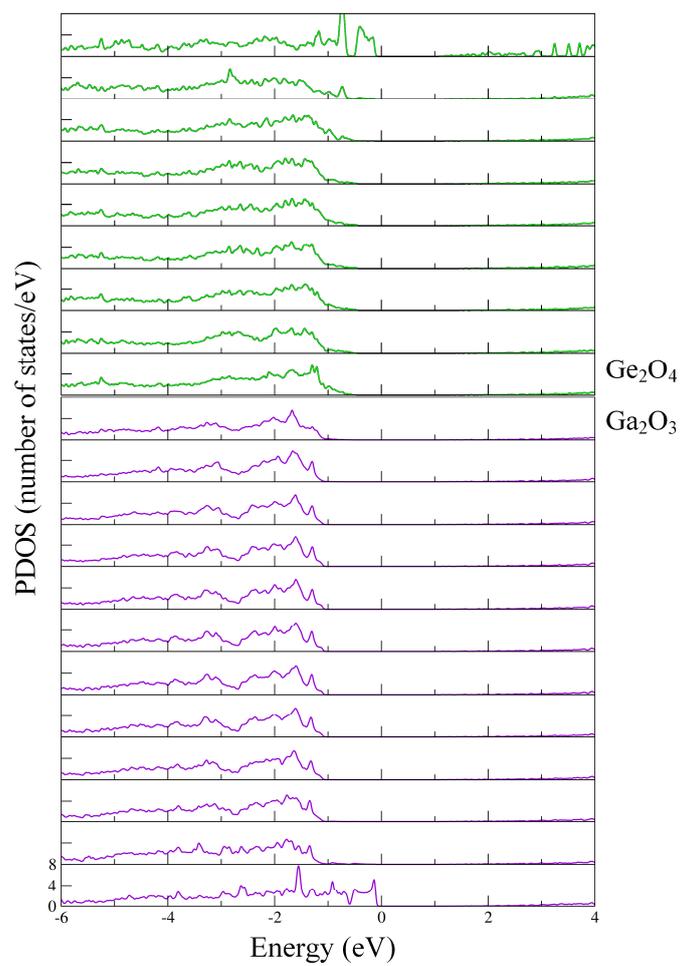

Fig. 2 Layer-by-layer projected density of states (PDOS) of the (100)A β-Ga$_2$O$_3$/(110) r-GeO$_2$ interface calculated with the PBEsol functional.

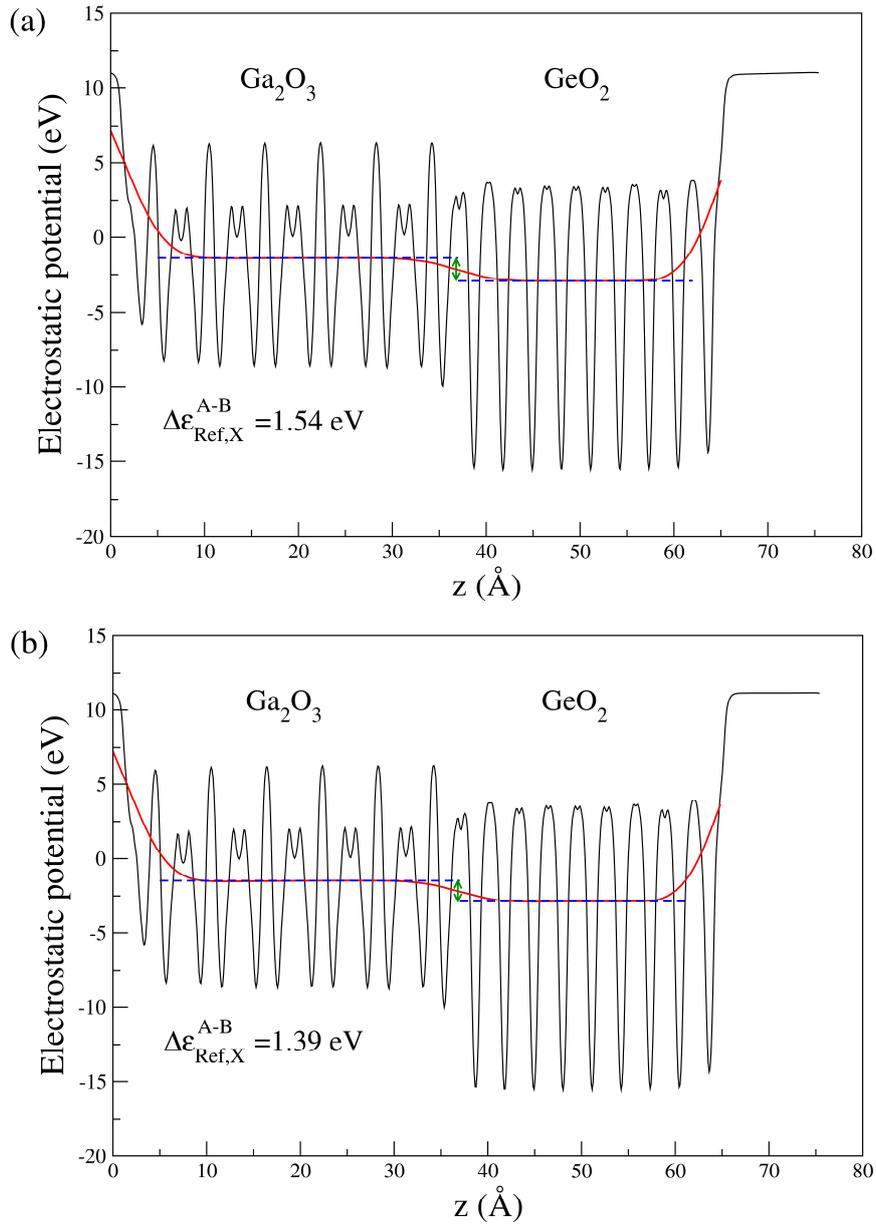

Fig. 3 Planar-averaged (black line) and macroscopically averaged (red line) electrostatic potentials along the c axis for the (100)A β-Ga$_2$O$_3$/(110) r-GeO$_2$ interface calculated using the PBEsol (a) and PBE0(α=0.2675) (b) functionals. The blue dashed lines are shown as a guide for the eye.

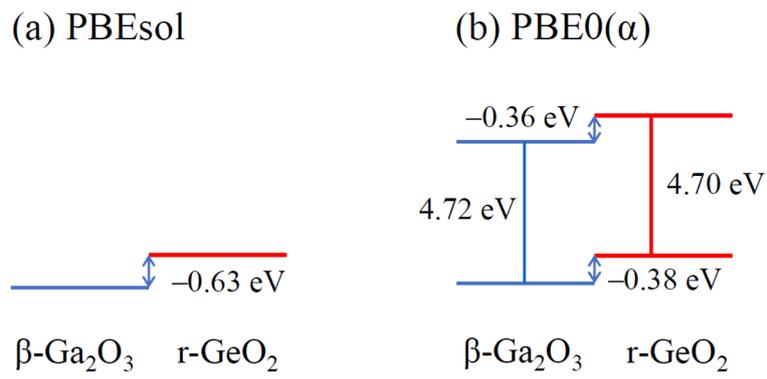

Fig. 4 Strained band offsets for the β-Ga₂O₃/r-GeO₂ heterojunction calculated using the PBEsol (a) and PBE0(α) functionals. Panel (b) also shows the bandgaps of strained bulks β-Ga₂O₃ and r-GeO₂.

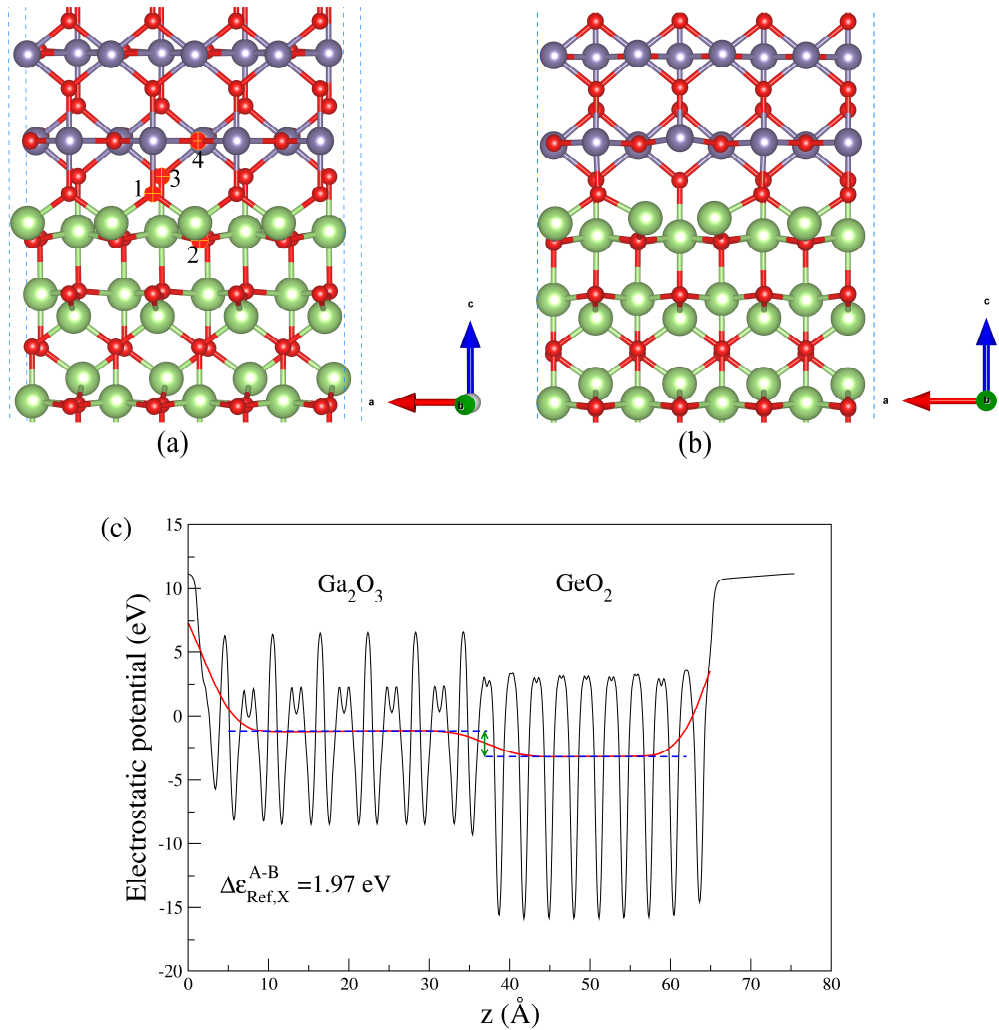

Fig. 5 (a) The sites of oxygen vacancy at the interface are indicated by numbers. (b) Relaxed atomic structure of an oxygen vacancy at site 1. (c) Reference level difference for the (100)A β-$Ga_2O_3$/(110) r-$GeO_2$ interface with an oxygen vacancy at site 1 calculated using the PBEsol functional.